# ON THE LORENTZ INVARIANT GRAVITATION FIELD THEORY

**Marije Ljolje**

Faculty of Electrical Engineering and Computing
University of Zagreb, Croatia
marije@tel.fer.hr

Summary- The theory of gravitation field within the special theory of relativity is analyzed.

# Introduction

The most accepted theory of gravitation is the one due to the Einstein where the gravitation effects are included in the space-time metric. Many aspects of this theory have been analyzed so far. But still there are serious open questions. For example R.Penrose writes:" *It is generally accepted that one of the most important of the unsolved problems of theoretical physics is to find the appropriate union of the generally relativity with quantum mechanics – in effect, to find a unified physics of the large and the small. It has been the conventional viewpoint on this issue that what is needed is the correct of the quantum gravity which, the most physicists, would mean the correct application of the standard rules of quantum (field) theory to Einstein's general relativity- or to some appropriate modification of that theory if it turns out that Einstein's theory resist all attempts at quantization in ordinary sense. My own viewpoint is significantly different…*" [1]. Similarly R.H.Kraichman writes: "*The general theory of relativity has long occupied a position of isolation with respect to the rest of contemporary fundamental physics. The special theory of relativity has been intimately and indispensably amalgamate with quantum mechanics in evolving current theoretical representation of elementary processes, but general theory, despite the elegance of its concepts, has not exhibited any real relation whatever to quantum physics. The attempts which have been made to connect general relativity and quantum mechanics have been directed largely at showing to what degree the two disciplines may be compatible rather than seeking a basic interdependence.*"[2].

Remarkable efforts have been put on connection to the classical framework especially in mathematical respect [3-9]. But almost of them finished Einstein's view on space–time. Exceptions in some sense are those like the one due to F.J.Belifante [6] and similar. However, a closed concept is mostly missing.

In the article we want to show in what degree with respect to observations, if at all, the theory of gravitation can be constructed fully inside of special theory of relativity. For the sake of simplicity we restrict the analysis to particles as source of the gravitation. The analysis we perform in analogy to the electromagnetic field theory.

In chapter 1 the free field is defined. Interaction of the field with particles is considered in chapter 2. A two particle system, with one heavy particle is considered in chapter 3. Conclusions are given in chapter 4.

1. Free G-field

   Sources of gravitation are masses. In the special theory of relativity they are fully present in the energy momentum tensor which for free particles is given by

   $$T^{\alpha\beta} = \mu c u^\alpha u^\beta \frac{ds}{dt}, \quad u^\alpha = \frac{dx^\alpha}{ds} \tag{1.1}$$

   $$\mu = \sum_i m_i \delta(\vec{r} - \vec{r}_i), \tag{1.2}$$

   The energy momentum tensor is also present in the Einstein's theory of gravitation but as source of space time curvature. Here we take it as analogue to the charge four-current of electromagnetic field theory. Corresponding Lagrange's variables are component of a second order symmetrical tensor $\Phi^{\alpha\beta}$.

   In the linear theory of Lagrangian density of the free field is then

   $$L_{free} = K F_{\alpha\beta\gamma} F^{\alpha\beta\gamma} \tag{1.3}$$

   where*

   $$F_{\alpha\beta\gamma} = \partial_\alpha \Phi_{\beta\gamma} \tag{1.4}$$

   and $K$ is constant.

   The Lagrangian's equations, according to

   $$\frac{\partial L}{\partial \chi} - \partial_\eta \frac{\partial L}{\partial(\partial_\eta \chi)} = 0 \tag{1.5}$$

   are given by

   $$\partial_\eta F_{\eta\alpha\beta} = 0. \tag{1.6}$$

   or

   $$\partial_\eta \partial^\eta \Phi^{\alpha\beta} = 0, \text{ (wave equation)}. \tag{1.7}$$

---

* It is possible to take fully symmetrical tensor

## 2. Interaction of the G-field with particles

The Lagrangian interaction scalar may be constructed by construction of tensors $\Phi^{\alpha\beta}$ and $T^{\alpha\beta}$, and by product $\Phi T$, where

$$\Phi = \Phi^\alpha_\alpha = \Phi_{00} - \Phi_{11} - \Phi_{22} - \Phi_{33}, \tag{2.1}$$

$$T = T^\alpha_\alpha = \mu c \frac{ds}{dt}. \tag{2.2}$$

Thus, we define

$$L_{int} = a\Phi_{\alpha\beta}T^{\alpha\beta} + b\Phi T, \tag{2.3}$$

where $a$ and $b$ are constants.

### 2.1. Lagrange's field equations

According to (1.5)

$$\frac{\partial L}{\partial \Phi_{\alpha\beta}} = \begin{cases} 2aT^{\alpha\beta} &, \alpha \neq \beta \\ a(T^{00} + bT) &, \alpha = \beta = 0 \\ a(T^{ii} - bT) &, \alpha = \beta = i \end{cases} \tag{2.4}$$

$$\frac{\partial L}{\partial(\partial_\eta \Phi_{\alpha\beta})} = 2K \frac{\partial(\partial_\xi \Phi_{\zeta\rho})}{\partial(\partial_\eta \Phi_{\alpha\beta})} \partial^\xi \Phi^{\zeta\rho} \tag{2.5}$$

and

$$\frac{\partial L}{\partial(\partial_\eta \Phi_{\alpha\beta})} = \begin{cases} 4k(\partial^\eta \Phi^{\alpha\beta}), \alpha \neq \beta \\ 2k(\partial^\eta \Phi^{\alpha\beta}), \alpha = \beta, \end{cases} \tag{2.6}$$

it follows

$$\partial_\eta \partial^\eta \Phi^{\alpha\beta} = \frac{1}{2k} \begin{cases} aT^{\alpha\beta} &, \alpha \neq \beta \\ aT^{00} + bT, \alpha = \beta = 0 \\ aT^{ii} + bT, \alpha = \beta = i \end{cases} \tag{2.7}$$

Retarded particular solution is given by

$$\Phi_{ret}^{\alpha\beta} = \frac{1}{8\pi k}\int\frac{d\vec{r}'}{|\vec{r}-\vec{r}'|}\begin{cases} aT^{\alpha\beta}(\vec{r}',t'), & \alpha\neq\beta \\ aT^{00}(\vec{r}',t')+bT(\vec{r}',t'), & \alpha=\beta=0 \\ aT^{ii}(\vec{r}',t')-bT(\vec{r}',t'), & \alpha=\beta=i \end{cases} \quad (2.8)$$

## 2.2. Lagrange's particle equations

The action integral of whole system is

$$S = \sum_i(-m_i c)\int ds_i + \frac{K}{3}\int\Phi_{\alpha\beta\gamma}\Phi^{\alpha\beta\gamma}d^3xdt + \int\left(a\Phi_{\alpha\beta}T^{\alpha\beta}+b\Phi T\right)d^3xdt. \quad (2.9)$$

It contains also self-interaction. The self-interaction may be reduced by the method of the classical electromagnetic field theory [10]. In this article we are not concerned with self-interaction. So, we simply omit it.

After elimination of self-interaction the variation of $S$ with respect to particle coordinate yields the particle equations motion. In the next section we consider a special system and then write them explicitly.

## 3. Two particle (without of self-interaction)

Let us consider a system of two particles interacting with G field with masses $m$ and $M$. Then one gets

$$T^{\alpha\beta} = mc\delta(\vec{r}-\vec{r}_m)u_m^\alpha u_m^\beta\frac{ds_m}{dt} + Mc\delta(\vec{r}-\vec{r}_M)u_M^\alpha u_M^\beta\frac{ds_M}{dt}, \quad (3.1)$$

$$T = mc\delta(\vec{r}-\vec{r}_m)\frac{ds_m}{dt} + Mc\delta(\vec{r}-\vec{r}_M)\frac{ds_M}{dt}, \quad (3.2)$$

$$\Phi_{ret}^{\alpha\beta} = \Phi_{ret\ m}^{\alpha\beta} + \Phi_{ret\ M}^{\alpha\beta}. \quad (3.3)$$

Excluding the self-interaction, the action integral becomes

$$S = -mc\int ds_m - Mc\int ds_M + \frac{1}{3}K\int\Phi_{\alpha\beta\gamma}\Phi^{\alpha\beta\gamma}d^3xdt + \int\left(a\Phi_{\alpha\beta M}T_m^{\alpha\beta}+b\Phi_M T_m\right)d^3xdt + \int\left(a\Phi_{\alpha\beta m}T_M^{\alpha\beta}+b\Phi_m T_M\right)d^3xdt. \quad (3.4)$$

Let be

$$M \gg m$$

so that we may take the heavy particle at rest in the origin of coordinate system. The action for the particle $m$ becomes

$$S_m = -mc\int ds_m + \int \left(a\Phi_{00M} T_m^{00} + b\Phi_{00M} T_m^{00}\right) d^3x dt. \qquad (3.5)$$

After substitution $\Phi_{00M}$ form (2.7) and $T_m^{\alpha\beta}$ form (1.1) it follows

$$S = -mc\int ds + \int \frac{1}{8\pi k} \frac{Mc^2}{r}(a+b)\left(au^{0^2} + b\right) mc \cdot ds, \qquad (3.6)$$

where we have omitted the subscript $m$. From here one gets the particle Lagrangian

$$L = \left[-mc^2 + \frac{1}{8\pi k} \frac{Mc^2}{r}(a+b)\left(au^{0^2} + b\right)\right]\sqrt{1 - \frac{v^2}{c^2}}. \qquad (3.7)$$

In the limit $v \ll c$ it becomes

$$L_{v \ll c} = -mc^2 + \frac{mv^2}{2} + \frac{Mm}{r}\frac{(a+b)^2 c^4}{8\pi k}. \qquad (3.8)$$

Compaction with Newton theory yields

$$\frac{(a+b)^2 c^4}{8\pi k} = G. \qquad (3.9)$$

It is one condition on constants $a$ and $b$.

By making use of (3.9) the Lagrangian (3.7) becomes

$$L = \left[-mc^2 + G\frac{Mm}{r}\frac{\left(au^{0^2} + b\right)}{(a+b)}\right]\sqrt{1 - \frac{v^2}{c^2}}. \qquad (3.10)$$

The corresponding Lagrange's equation are given by

$$\frac{\partial L}{\partial \vec{r}} - \frac{d}{dt}\frac{\partial L}{\partial \vec{v}} = 0. \qquad (3.11)$$

We evaluate this equation for $v \ll c$ and up to the terms of the order $v^2/c^2$. This approximation yields

$$L = mc^2\left[-1 + \frac{v^2}{2c^2} + \frac{GM}{c^2 r}\left(1 + f\frac{v^2}{c^2}\right)\right], \qquad (3.12)$$

$$f = \frac{a-b}{2(a+b)}, \qquad (3.13)$$

$$\frac{\partial L}{\partial \vec{v}} = mGM\left(1 + f\frac{v^2}{c^2}\right)\nabla\frac{1}{r}, \qquad (3.14)$$

$$\frac{\partial L}{\partial \vec{v}} = m\vec{v}\left(1 + 2f\frac{GM}{rc^2}\right), \qquad (3.15)$$

and

$$\frac{d}{dt}(\vec{v}F) = GM\left(1 + f\frac{v^2}{c^2}\right)\nabla\frac{1}{r}, \qquad (3.16)$$

$$F = 1 + 2f\frac{GM}{rc^2}. \qquad (3.17)$$

By marking use of cylindrical coordinates

$$x = r\cos\varphi, \quad y = r\sin\varphi, \quad z = z$$

and taking $z = 0$ one gets

$$\frac{d}{dt}(\dot{r}F) - r\dot{\varphi}^2 F + \left(1 + f\frac{v^2}{c^2}\right)\frac{GM}{r} = 0, \qquad (3.18)$$

$$\frac{d}{dt}(r^2\dot{\varphi}F) = 0. \qquad (3.19)$$

Equation (3.19) yields

$$r^2\dot{\varphi}F = h \text{ (constant)} \qquad (3.20)$$

Introduction

$$r = \frac{1}{u}, \quad u(\varphi), \qquad (3.21)$$

yields

$$F = 1 + 2f\frac{GM}{rc^2}u, \qquad (3.22)$$

$$v^2 = \dot{r}^2 + r^2\dot{\varphi}^2 = \frac{h^2}{F^2}(u^2 + u'^2), \quad u' = \frac{du}{d\varphi}, \tag{3.23}$$

and

$$u'' + \left(1 - 2f\frac{G^2M^2}{c^2h^2}\right)u - \frac{MG}{h^2} - \frac{MG}{c^2}f(u^2 + u'^2) = 0. \tag{3.24}$$

Application of the perturbation method gives

$$u = u^{(0)} + u^{(1)} + \ldots, \tag{3.25}$$

$$u^{(0)''} + u^{(1)} - \frac{MG}{h^2} = 0, \tag{3.26}$$

$$u^{(0)''} + u^{(1)} - \left[2f\frac{M^2G^2}{c^2h^2}u^{(0)} + f\frac{MG}{c^2}\left(u^{(0)2} + u^{(0)'2}\right)\right] = 0, \tag{3.27}$$

$$\vdots$$

$$u^{(0)} = \frac{MG}{c^2} + D\cos\varphi, \tag{3.28}$$

$$u^{(1)} = \left[\frac{MG}{c^2}f\left(2\frac{M^2G^2}{h^4} + D^2\right) + 2\frac{M^2G^2}{c^2h^2}fD\varphi\sin\varphi\right] = 0 \tag{3.29}$$

and

$$u^{(0)} + u^{(1)} =$$
$$= \frac{MG}{c^2}\left[1 + f\left(2\frac{M^2G^2}{h^4} + D^2\right)\right] + D\left(\cos\varphi + 2\frac{M^2G^2}{c^2h^2}f\varphi\sin\varphi\right). \tag{3.30}$$

Due to $\frac{2M^2G^2}{c^2h^2}f \ll 1$ it may be written in the form

$$u^{(0)} + u^{(1)} =$$
$$= \frac{MG}{c^2}\left[1 + f\left(2\frac{M^2G^2}{h^4} + D^2\right)\right] + D\cos\left(\varphi - 2\frac{M^2G^2}{c^2h^2}f\varphi\right). \tag{3.31}$$

The second term shows perihelion shift. It is given by

$$\left(1 - 2\frac{M^2G^2}{c^2h^2}f\right)(2\pi + \delta\varphi) = 2\pi \tag{3.32}$$

and form here

$$\delta\varphi = \frac{4\pi f \dfrac{M^2 G^2}{c^2 h^2}}{1 - 2f \dfrac{M^2 G^2}{c^2 h^2}} \doteq 4\pi f \frac{M^2 G^2}{c^2 h^2} \qquad (3.33)$$

The parameters *a* and *b* are subject to the condition (3.9). The second condition may be requirement that $\delta\varphi$ agrees with observations. For $f = 3/2,\ b = -a/2^*$, $\delta\varphi^*$, given by (3.33) agrees with that one due to Einstein's.

## 4. Conclusion

It is possible to construct the gravitation field theory within the special theory of relativity in accordance to particle observations. That indicates justification of investigations the gravitation field within the theory of special relativity before going further

---

* Then $a = 2/c^2,\ K = 1/8\pi G$

References

[1]  R.Penrose, Chaos, Solitons Fractals, Vo.10, No.2-3, pp 581-611 (1999),
[2]  R.H.Kraichan, Phys. Rev. 98, 1118 (1955),
[3]  P.Havas and J.N.Goldberg, Phys. Rev. 128, 398 (1962),
[4]  M.Mathisson, Acta Phys. Polon. 6, 163 (1937),
[5]  S.N.Gupta, Proc. Phys. Soc. (London) A65, 608 (1952),
[6]  F.J.Belifante, Phys. Rev. 89, 914 (1953),
[7]  B.Bertotti, Nuovo cimento 4, 898 (1956),
[8]  D. Gessler, Z. Naturforsch. 14a, 689 (1959),
[9]  R.P.Kerr, Nuovo cimento 13, 469,492 and 693 (1959),
[10] S.Botrić and K.Ljolje, I1 Nuovo cimento B, 107, 51 (1992).